\RequirePackage[2020-02-02]{latexrelease}
\documentclass
[aps,preprint,showkeys,a4paper,tightenlines,superscriptaddress]{revtex4}%
\usepackage{eurosym}
\usepackage{amsfonts}
\usepackage{amsmath}
\usepackage{amssymb}
\usepackage{graphicx}
\usepackage{placeins}
\usepackage{subfig}
\usepackage{caption}
\usepackage{xcolor}%
\providecommand{\U}[1]{\protect\rule{.1in}{.1in}}
\providecommand{\U}[1]{\protect\rule{.1in}{.1in}}

\begin{document}
\title{Enhancement of adhesion strength in viscoelastic unsteady contacts}
\author{C. Mandriota}
\affiliation{Department of Mechanics, Mathematics and Management, Politecnico of Bari, Via Orabona, 4, 70125, Bari, Italy}
\author{N. Menga}
\affiliation{Department of Mechanics, Mathematics and Management, Politecnico of Bari, Via Orabona, 4, 70125, Bari, Italy}
\affiliation{Corresponding author. Email: nicola.menga@poliba.it}
\author{G. Carbone}
\affiliation{Department of Mechanics, Mathematics and Management, Politecnico of Bari, Via Orabona, 4, 70125, Bari, Italy}
\affiliation{CNR - Institute for Photonics and Nanotechnologies U.O.S. Bari, Physics
Department \textquotedblright M. Merlin\textquotedblright, via Amendola 173,
70126 Bari, Italy}

\begin{abstract}
We present a general energy approach to study the unsteady adhesive contact of
viscoelastic materials. Under the assumption of infinitely short-range
adhesive interactions, we exploit the principle of virtual work to generalize
Griffith's local energy balance at contact edges to the case of a
non-conservative (viscoelastic) material, subjected to a generic contact
time-history. We apply the proposed energy balance criterion to study the
approach-retraction motion of a rigid sphere in contact with a viscoelastic
half-space. A strong interplay between adhesion and viscoelastic hysteretic
losses is reported which can lead to strong increased adhesion strength, depending
on the loading history. Specifically, two different mechanisms are found to govern the increase of pull-off force during either approach -- retraction cycles and approach -- full
relaxation -- retraction tests. In the former case, hysteretic losses
occurring close to the circular perimeter of the contact play a major role,
significantly enhancing the energy release rate. In the latter case, instead, the pull-off enhancement mostly depends on the
glassy response of the whole (bulk) material which, triggered by the fast retraction
after relaxation, leads to a sort of `frozen' state and results in a flat-punch-like
detachment mechanism (i.e., constant contact area). In this case, the JKR
theory of adhesive contact cannot be invoked to relate the observed pull-off
force to the effective adhesion energy, i.e. the energy release rate $G$, and
strongly overestimates it. Therefore, a rigorous mathematical procedure is
also proposed to correctly calculate the energy release rate in viscoelastic
dissipative contacts.
\end{abstract}
\maketitle

\section{Introduction}

The Johnson, Kendall and Roberts (JKR) seminal study \cite{jkr} on adhesive
elastic spheres is a milestone in the field of adhesive contact mechanics of
soft elastic solids. Moving from the assumption of infinitely short-range
adhesive interactions and exploiting the Griffith energy balance criterion
\cite{Griffith}, the JKR\ model was able to provide results in perfect
agreement with experiments thus paving the way for modelling a wide class of
elastic adhesive smooth \cite{Carbone2004,Guduru2007} and rough contacts
\cite{Greenwood,CarboneScaraggi,CarboneMangialardiPersson}, as well as other
applications such as the peeling of thin tapes
\cite{Menga2018peeling,Menga2020peeling}.\ By following the same energy
balance principle, the influence of tangential stresses in sliding adhesive
contacts can be investigated\cite{Menga2018,Menga2018C}, as well as the effect
of thickness of the elastic coatings \cite{Menga2019layered,Stan}, and of
different geometries and boundary conditions
\cite{Carbone2008,Menga2016,Lin2008,Carbone2011}. As long as the Tabor
parameter is sufficiently large \cite{Maugis1992}, the energy balance approach
has the advantage to correctly model adhesive contacts only requiring a few
quantities as inputs: the Young modulus and Poisson ratio of the isotropic
elastic material, and the adhesion energy per unit area $\Delta\gamma$. The
latter, in particular, has a very simple definition in terms of macroscopic
quantities and can be easily measured through reliable experimental
procedures, regardless of the detailed interfacial gap dependence of the
molecular interactions. Hence, once the experimental data are acquired using,
for instance, micro-scale optical microscopy \cite{Gruhn}, scratch tests
\cite{Das}, peeling processes \cite{Raegen}, or macro-scale spherical
indentation-retraction tests \cite{Shull2002,Hui2015}, $\Delta\gamma$ can be
easily derived by fitting them with the corresponding elastic contact model.

Nonetheless, the JKR contact theory falls short in tackling adhesive contact
problems of rubber-like materials when quasi-static conditions are not
ensured, as clearly indicated by several experimental evidences. This is often
ascribed to the intrinsic viscoelastic response of rubber, whose
non-conservative nature makes Griffith's energy balance no longer valid.
Indeed, a velocity-dependent response is experimentally observed, e.g. in
rolling contacts of natural rubber\ \cite{charmet} with respect to the real
adhesion force. Similarly, JKR-like dynamic tests on siloxane and acrylic
elastomers reveal velocity-dependent hysteresis, for both approach-retraction
\cite{Ahn,Das2021} and oscillating \cite{Charrault} tests, with highly
enhanced effective adhesion energy reported during retraction. Experiments
have also been devoted to investigate the effect of dwell time before
retraction \cite{davis,luengo}, preload \cite{Baek,Li}, and micro vibration
during retraction \cite{Shui,wahl,Ebenstein}. Although some attempts have been made to predict the retraction
behavior by introducing velocity-dependent (usually, power law) adhesion
energy terms \cite{Ahn,Deruelle,Violano,Lorenz2013}, the overall effect of
viscoelasticity in adhesive contacts is still not fully understood. Indeed,
the interplay between viscoelasticity and adhesion in unsteady contacts is a
complex phenomenon, affected by the overall loading history and not only by
the instantaneous crack speed at the contact boundary, as demonstrated by
several experimental studies highlighting the effect of the dwell time before
retraction \cite{davis,luengo}, the preload \cite{Baek,Li}, and the imposed
micro vibration during retraction \cite{Shui,wahl,Ebenstein}. More
importantly, variational approaches to viscoelastic continuum mechanics
\cite{Schapery1964,Biot1955,Gurtin} usually result in greater complexity
compared to corresponding elastic cases, therefore discouraging the attempts
to adapt JKR approach to viscoelastic contacts. As a consequence, the common
approach to investigate adhesive viscoelastic contacts is to assume that
viscoelastic losses occur very locally at the boundary of the contact, while
in the bulk the material response is governed by the soft elastic modulus
\cite{Greenwood2006,Johnson2000,Ahn,Lorenz2013}. This assumption, known as the
`small-scale viscoelasticity' hypothesis, has been shown to be very effective
and accurate for systems which do not present any specific length scale, e.g.
a semi-infinite crack propagating in infinite viscoelastic sheet, or for
contact problems where the characteristic size $a$ of the contact is $a\gg
V\tau$ where $V$ is a characteristic velocity and $\tau$ the relaxation time
of the material. However, when the above conditions are not fulfilled, recent
studies \cite{Carbone2022,Mandriota} have proved that the small-scale
viscoelasticity assumption leads to significant qualitative and quantitative
errors, which are mainly related to the neglected bulk viscoelastic
dissipation. In the attempt to deal also with these cases, most of the
existing studies exploit local force equilibrium where the contacting surface
are discretized in particles or elements which interact with the corresponding
particles/elements of the counter surface through local forces, derived by
gap-dependent potentials, e.g. $m-n$ potentials \cite{Mueser2022pot}, as the
Lennard-Jones (LJ) law
\cite{Luo2024,Afferrante2023,Afferrante2022,Violano2022,Jiang}, exponentially
short-range laws \cite{Mueser2022,Persson2014}, and cohesive-zone models
specifically designed for contact \cite{Lin,Haiat,Haiat2007,Perez2023} and
fracture mechanics \cite{Schapery1989,Schapery1975} problems. However, as
clearly observed by Persson \cite{Persson2021} and Greenwood
\cite{Greenwood2007}, the contact behavior is weakly affected by the specific
law implemented to describe the interfacial adhesive interaction, provided
that the range of microscopic gap dependent law is much shorter than any other
length scale involved in the problem. This suggests that most of the phenomena
in viscoelastic contacts can be captured by general energy equilibrium, as
indeed shown at least for the case steady-state viscoelastic sliding contacts
in Ref. \cite{Carbone2022,Mandriota}.

The present study generalizes the adhesive viscoelastic contact theory
presented in \cite{Carbone2022,Mandriota} to the case of unsteady conditions
(dynamic loading), thus formally deriving a general Griffith-like criterion
for viscoelastic contacts and unsteady crack propagation in hysteretic
materials (e.g. delayed fracture of soft polymers). We derive the energy
balance by relying on the D'Alembert principle of virtual works: the variation
of adhesion energy due to a virtual change of the contact area (the Lagrangian
coordinate) must be precisely balanced by the virtual work of internal
stresses. Our findings are in perfect agreement with aforementioned
experimental results and LJ based numerical calculations
\cite{Afferrante2022,Violano2022,Lin2002,Jiang}, and clearly indicate that
viscoelasticity plays a major role in affecting adhesion enhancement depending
on the specific loading-history. Our theoretical approach also provides very
profound insights into the physical mechanisms governing experimentally
observed phenomena such as the enhancement of pull-off force during fast
retraction \cite{Giri,Li,Linghu,Ahn}, the hysteresis during the
approach-retraction cycle \cite{Ahn,Das2021}, and effective contact stiffness
during high frequency oscillations \cite{wahl,Ebenstein}. Overall, the present
theory may be of interest to quantify the effects of adhesion in several
engineering applications involving viscoelastic polymeric materials such as,
for instance, structural adhesives \cite{Sancaktar,Bitner}, pressure-sensitive
adhesives \cite{Chang}, protective coatings \cite{Montazeri}, bioinspired
adhesives \cite{Favi,Purtov}, orthopedic applications \cite{Shah,Zaokari},
micro-electro-mechanical systems \cite{Ding,Taran}, micro-manipulations and
micro-assembly \cite{Cecil,Alogla}.

\section{Formulation}

\begin{figure}[ptbh]
\includegraphics[width=.85\textwidth]{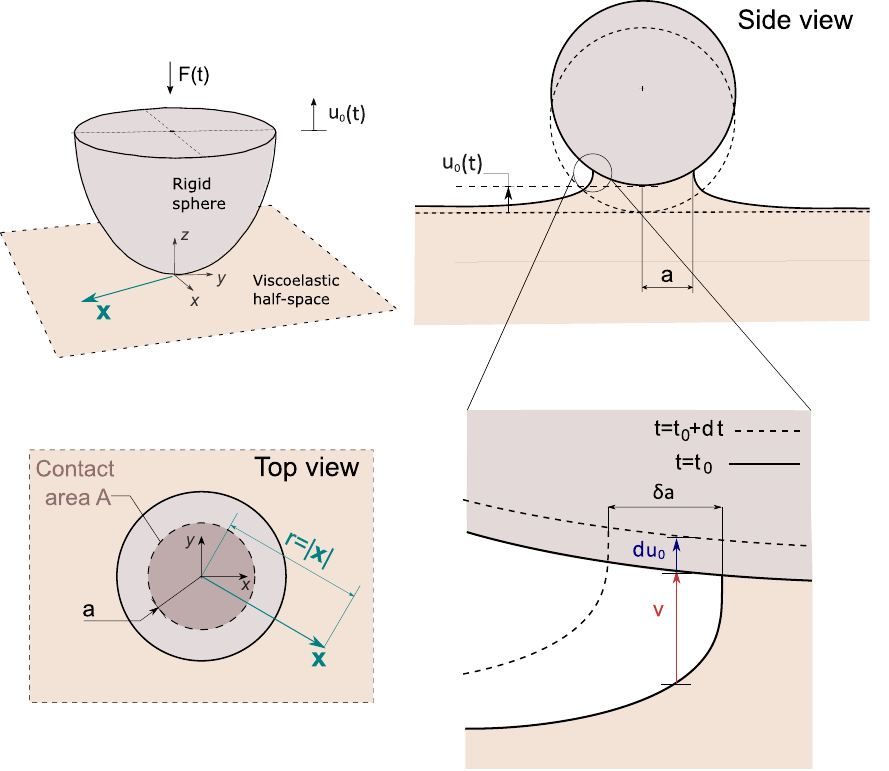}\caption{The schematic of the
adhesive contact between a viscoelastic half-space and a rigid sphere, with
time-varying normal rigid displacement $u_{0}(t)$. The inset represents the
virtual component $v$ of the local displacement $v+du_{0}$ close to the
contact edge associated with the contact area variation $\delta a$ and
indenter rigid displacement $du_{0}$.}%
\label{fig1}%
\end{figure}

We consider the adhesive contact between a linear viscoelastic half-space and
a rigid sphere of radius $R$ subjected to a time-varying rigid normal
displacement $u_{0}(t)$, as shown in Fig \ref{fig1}. Following \cite{Carbone2013,Putignano2022}, the normal
displacement field $u(\mathbf{x},t)$ of the viscoelastic half-space surface is
given by
\begin{equation}
u(\mathbf{x},t)=J(0)\int dx_{1}^{2}\mathcal{G}(\mathbf{x-x}_{1})\sigma
(\mathbf{x}_{1},t)+\int_{-\infty}^{t}dt_{1}\dot{J}(t-t_{1})\int dx_{1}%
^{2}\mathcal{G}(\mathbf{x}-\mathbf{x}_{1})\sigma(\mathbf{x}_{1},t_{1})
\label{convolution}%
\end{equation}
where $\mathbf{x}$ is the in-plane position vector, $t$ is the time variable,
$\sigma(\mathbf{x},t)$ is the interfacial normal stress distribution, $J(t)$
is the viscoelastic creep function, and $\mathcal{G}(\mathbf{x})=(1-\nu
^{2})|\mathbf{x}|^{-1}/\pi$. As we are considering the approach/retraction of
a sphere, the problem at hand, which belong to the class of mixed value
problems, is axisymmetric, i.e., at each time step $t$ all quantities depend
only on the distance $r=|\mathbf{x}|$ from the contact center (see Fig.
\ref{fig1}). Under the assumption of infinitely short range interfacial
adhesive interactions, outside the circular contact region of radius $a(t)$
the surface stresses must vanish, i.e. $\sigma(r,t)=0$ for $r>a(t)$, whilst
within the contact region the surface displacement field is prescribed, i.e.
$u(r,t)=s(r,t)$ for $r\leq a(t)$, with $s(r,t)=u_{0}(t)+r^{2}/(2R)$ being the
spherical indenter surface at time $t$. We also define the local interfacial
gap as $g\left(  r,t\right)  =s\left(  r,t\right)  -u\left(  r,t\right)  $, so
that $g\left(  r,t\right)  =0$ for $r\leq a(t)$. Note that, given a value of
the rigid normal displacement $u_{0}(t)$, Eq. (\ref{convolution}) cannot be
solved as the contact radius $a(t)$ is not known, and an additional closure
equation is needed to solve the problem. In the case of adhesiveless contacts,
the closure condition is simply that the interfacial contact stress must
vanish at the boundary of the contact area; however, for adhesive contacts,
tensile stress can also be sustained, and an energy based condition should be
identified. In virtue of the principle of virtual works (PVW), the equilibrium
configuration at time $t$ requires that the work of external forces $\delta
L_{E}$ due to an admissible virtual displacements field equates the work of
internal stresses $\delta L_{I}$ due to the corresponding compatible virtual
strain field. Since the virtual displacements field $\delta\mathbf{v}(r,z,t)$
must only satisfy the boundary conditions at time $t$ (i.e., the kinematic
constraints), we have $\delta\mathbf{v}(x,z,t)=0$ within the contact region,
i.e. where the time-dependent constraint $u\left(  r,t\right)  =s\left(
r,t\right)  $ is prescribed. Then, neglecting body forces, at equilibrium we
have
\begin{equation}
\delta L_{I}=\int_{W}\sigma_{ij}\delta\varepsilon_{ij}dV=\int_{\partial
W}\mathbf{\sigma}\cdot\delta\mathbf{v}dA=\delta L_{E} \label{PVW}%
\end{equation}
for any admissible virtual displacement $\delta\mathbf{v}$ and its associated
internal strain tensor $\delta\varepsilon_{ij}$, where $\mathbf{\sigma}$ is
the surface stress field, and $\sigma_{ij}$ is the internal stress tensor.

For the contact problem represented in Fig. \ref{fig1}, we assume a virtual
variation (i.e., at fixed time $t$) of the contact configuration so that the
contact radius increases from $a\left(  t\right)  $ to $a\left(  t\right)
+\delta a$. Consequently, the asymptotic surface displacements at the contact
edge [i.e., for $\left\vert r-a\left(  t\right)  \right\vert \ll a\left(
t\right)  $] changes by the quantity $v^{-}\left(  r\right)  =g\left[  r\geq
a\left(  t\right)  ,t\right]  $. This can be described by the virtual
displacement process $v(r,\eta)=v^{-}(r)H(\eta)$, where $H(\eta)$ is the unit
step function, and $\eta$ is the process parameter spanning the entire real
axis. Therefore, at each step of the process, the virtual normal displacement
$\delta v(r,\eta)$ obeys the equation%
\begin{equation}
\delta v(r,\eta)=\frac{\partial v}{\partial\eta}d\eta=v^{-}(r)\delta
(\eta)d\eta\label{crack virtual displacement}%
\end{equation}
where $\delta(\eta)$ is the Dirac delta function. Similarly the asymptotic
stress distribution $\sigma_{\mathrm{a}}$ close to the boundary of the contact
area [i.e., for $\left\vert a\left(  t\right)  -r\right\vert \ll a\left(
t\right)  $] has the form $\sigma_{\mathrm{a}}\left[  a\left(  t\right)
-r,t\right]  =\sigma\left[  r<a\left(  t\right)  ,t\right]  $. Therefore,
during the virtual displacement process governed by the parameter $\eta$, the
corresponding asymptotic surface stresses are given by%
\begin{equation}
\sigma(r,\eta)=\sigma^{+}(r)H(\eta) \label{surface stress}%
\end{equation}
where $\sigma^{+}(r)=\sigma_{\mathrm{a}}\left[  a\left(  t\right)  +\delta
a-r,t\right]  $. It follows that during the entire $\eta$-governed process,
the total virtual work $\left(  \delta L_{I}\right)  _{\mathrm{T}}$ of
internal stresses due to the contact radius virtual variation from $a\left(
t\right)  $ to $a\left(  t\right)  +\delta a$ can be calculated as
\begin{equation}
\left(  \delta L_{I}\right)  _{\mathrm{T}}=2\pi\int_{a}^{a+\delta a}%
rdr\int_{-\infty}^{\infty}d\eta\sigma(r,\eta)\frac{\partial v}{\partial\eta
}=2\pi\int_{a}^{a+\delta a}rdr\sigma^{+}\left(  r\right)  v^{-}\left(
r\right)  \int_{-\infty}^{\infty}d\eta H\left(  \eta\right)  \delta\left(
\eta\right)  \label{balancing equations}%
\end{equation}
and, recalling that $\int_{-\infty}^{\infty}d\eta H\left(  \eta\right)
\delta\left(  \eta\right)  =1/2$, we finally get
\begin{equation}
\left(  \delta L_{I}\right)  _{\mathrm{T}}=\pi a\int_{a}^{a+\delta a}%
dr\sigma^{+}\left(  r\right)  v^{-}\left(  r\right)
\label{virtual work of internal stresses}%
\end{equation}
The virtual (external) work of adhesive forces during the entire displacement
process is instead%
\begin{equation}
\left(  \delta L_{E}\right)  _{\mathrm{T}}=2\pi\Delta\gamma a\delta a
\label{work of adhesive forces}%
\end{equation}
Thus, exploiting Eq. (\ref{PVW}) at each single step of the displacement
process, the energy balance gives
\begin{equation}
\left(  \delta L_{I}\right)  _{\mathrm{T}}=\left(  \delta L_{E}\right)
_{\mathrm{T}} \label{total virtual work}%
\end{equation}
which, using Eqs. (\ref{virtual work of internal stresses}%
,\ref{work of adhesive forces}), can be rewritten as%
\begin{equation}
\frac{1}{2\delta a}\int_{a}^{a+\delta a}dr\sigma^{+}(r)v^{-}(r)=\Delta
\gamma\label{closure}%
\end{equation}
Eq. (\ref{closure}) represents the generalization of the Griffith fracture
criterion for unsteady contacts and holds true for both elastic and viscoelastic materials. The positive quantity $\delta a$ should be chosen of
the same order of magnitude of the the so-called `process zone' at the contact
edges \cite{de Gennes}. Specific cases, such as thin pressure-sensitive membrane \cite{Chopin,Creton}, might require replacing in Eq. (\ref{closure}) the adhesion energy $\Delta\gamma$ with a modified energy of adhesion which depends on both the propagation speed of the crack tip and temperature of the process zone. The quantity $\delta a$ or equivalently the length of
the process zone is an additional (short) length scale, whose choice does not
affects the physical qualitative behavior of the viscoelastic contact problem
at hand, as it only shifts the frequency of the local excitation occurring
close to the boundary of the contact. Notably, Eq. \ref{closure} holds true
also for the case of steady state viscoelastic adhesive contact and is
formally the same as that derived in \cite{Carbone2022,Mandriota} following a
different argument.

\section{Results and discussion\label{flat punch}}

We consider a viscoelastic material with a single relaxation-time $\tau$ and
relaxation function given by%
\begin{equation}
G(t)=\mathcal{H}(t)\left\{  E_{\infty}+\left(  E_{0}-E_{\infty}\right)  \left[
1-\exp\left(  -t/\tau\right)  \right]  \right\}  \label{relaxation}%
\end{equation}
also corresponding to the creep function given by%
\begin{equation}
J(t)=\mathcal{H}(t)\left\{  \frac{1}{E_{\infty}}+\left(  \frac{1}{E_{0}}%
-\frac{1}{E_{\infty}}\right)  \left[  1-\exp\left(  -\frac{E_{0}}{E_{\infty}%
}\frac{t}{\tau}\right)  \right]  \right\}  \label{creep2}%
\end{equation}
where $E_{0}$, and $E_{\infty}$ are the low-frequency and very high-frequency
viscoelastic moduli of the material respectively. Real rubber-like materials usually present more complex rheology with several relaxation times, so that a Prony series description of the relaxation-creep functions is required; however, qualitative physical insights can still be drawn with a single-relaxation time.

Results are shown in terms of the following dimensionless quantities:
$\tilde{p}=\left(  1-\nu^{2}\right)  p/\left(  \pi E_{0}\right)  $, $\tilde
{u}=u/R$, $\tilde{\gamma}=\left(  1-\nu^{2}\right)  \Delta\gamma/\left(  \pi
E_{0}R\right)  $, $\tilde{F}=\left(  1-\nu^{2}\right)  F/\left(  \pi
E_{0}R^{2}\right)  $, $\tilde{V}=V\tau/R$, $\tilde{u}_{0}=u_{0}/R$,
$\tilde{\Delta}=\Delta/R$, $\tilde{t}=t/\tau$, $\tilde{r}=r/R$, $\tilde
{a}=a/R$, with $p\left(  r,t\right)  =-\sigma\left(  r,t\right)  $ being the
contact pressure, $\Delta=-u_{0}$ the contact penetration, and $F=\int
d^{2}xp\left(  r,t\right)  $ the normal compressive applied force. In our
calculations, we set unless $\delta\tilde{a}=0.013$, $\tilde{\gamma}%
=1.6\times10^{-4}$ (except in Fig. \ref{fig6}), and $E_{\infty}/E_{0}=10$. We
also define the normal approach-retraction speed of the sphere as
$V=\left\vert \dot{u}_{0}\right\vert =\left\vert \dot{\Delta}\right\vert $,
where the superposed dot `$\cdot$' stands for the time derivative. Notice
that, unless differently specified, $V$ is the controlled parameter in our calculations.

\begin{figure}[ptbh]
\includegraphics[width=.98\textwidth]{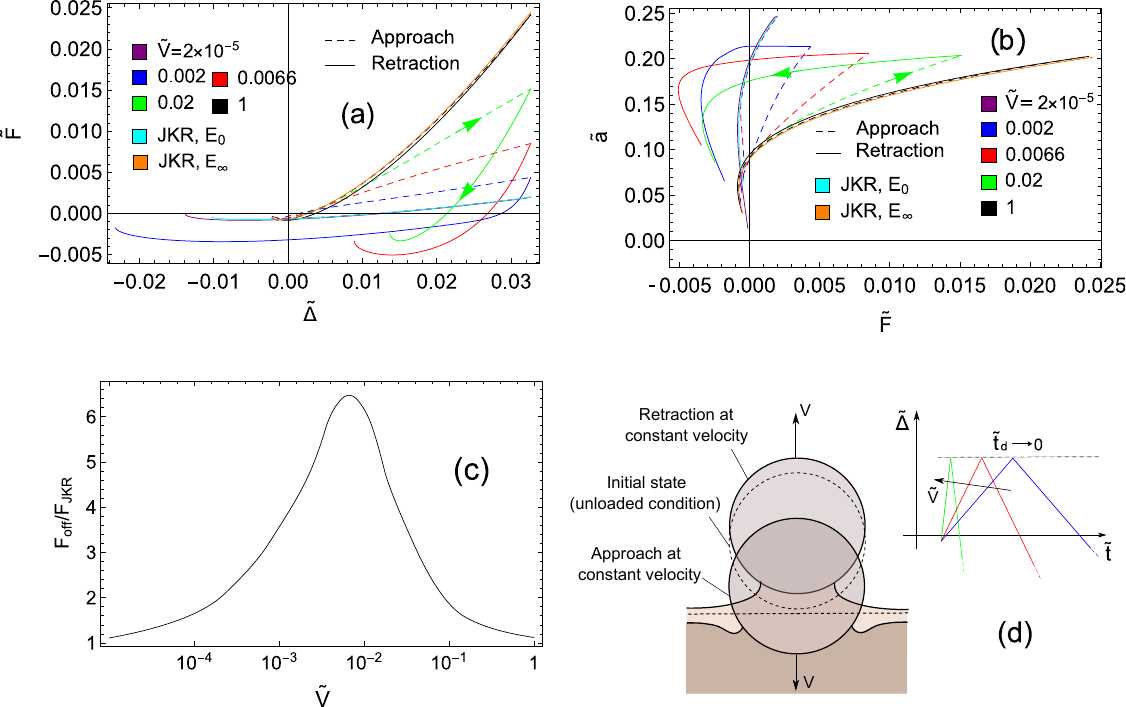}
\par
\caption{Approach-retraction cycles at different dimensionless sphere velocity
$\tilde{V}$. Dashed and solid lines refer, respectively, to indentation and
retraction, until pull-off occurs. (a) The dimensionless applied load
$\tilde{F}$ vs. the dimensionless indentation depth $\tilde{\Delta}$. (b) The
dimensionless contact radius $\tilde{a}$ vs. the dimensionless applied load
$\tilde{F}$. The JKR elastic curves corresponding to moduli $E_{0}$ and
$E_{\infty}$ are reported for comparison. (c) The viscoelastic-elastic pull-off force ratio
$F_{\text{\textrm{off}}}/F_{\text{\textrm{JKR}}}$ as function of the sphere velocity. (d) The process
schematic and the qualitative dimensionless penetration $\tilde{\Delta}$
time-history. Results are shown for $\tilde{\gamma}=1.6\times10^{-4},$
$E_{\infty}/E_{0}=10$.\newline}%
\label{fig2}%
\end{figure}

Firstly, we focus on the effect of the imposed indenter dimensionless speed
$\tilde{V}$ in approach--retraction (A-R) cycles with vanishing dwell time
$\Tilde{t}_{d}=t_{d}/\tau\rightarrow0$, as reported in Fig. \ref{fig2}.
Calculations are initialized assuming an instantaneous jump into contact at
$\tilde{\Delta}=0$. Therefore, regardless of $\tilde{V}$, the system response
is purely elastic and follows the adhesive JKR solution with the elastic
modulus given by $E_{\infty}$. At very low A-R speed (see results for
$\tilde{V}=2\times10^{-5}$), the loading process is sufficiently slow to allow
for full relaxation of the viscoelastic material, and the soft elastic JKR
response with low frequency modulus $E_{0}$ is recovered during both approach
and retraction (i.e., no hysteresis is observed). Similarly, at very high
speed (i.e., $\tilde{V}\gg1$), the material behaves as a stiff elastic body
and the glassy elastic JKR response occurs, with vanishing hysteresis during
the loading-unloading cycles. As expected, in both cases the maximum tensile
load (i.e. the pull-off force) takes the same value, independently of the
value of the effective elastic modulus. Nonetheless, results in Fig.
\ref{fig2} clearly show that viscoelastic dissipation induces large adhesive
hysteresis in A-R cycles at intermediate values of $\tilde{V}$. These findings
are in agreement with experimental results \cite{Giri,Baek,Li,Ahn} and show
that, during retraction, the system is able to withstand significantly larger
tensile loads compared to the elastic case, as also shown in
\cite{Violano,Li,Baek,Lorenz2013,Das2021}. A closer look at Fig. \ref{fig2}(a)
reveals that the maximum tensile stress (pull-off) can either occur at larger
retraction distances compared to the elastic JKR and, for relatively large A-R
speeds, even at positive penetration $\Delta>0$, in agreement with
experimental results \cite{Linghu,Giri}. In the latter case, while
approaching, the material has not yet reached the viscoelastic glassy
response. However, when the indenter motion is reversed (i.e., the speed jumps
from $\dot{\Delta}=V$ to $\dot{\Delta}=-V$), the glassy behavior is triggered
(with elastic modulus $E_{\infty}$) and, since retraction occurs sufficiently
fast, viscoelastic dissipation prevents the material from relaxing and
detachment occurs at positive values of penetration $\Delta$, with contact
area and tensile load much larger than the elastic case.

\begin{figure}[ptbh]
\includegraphics[width=.8\textwidth]{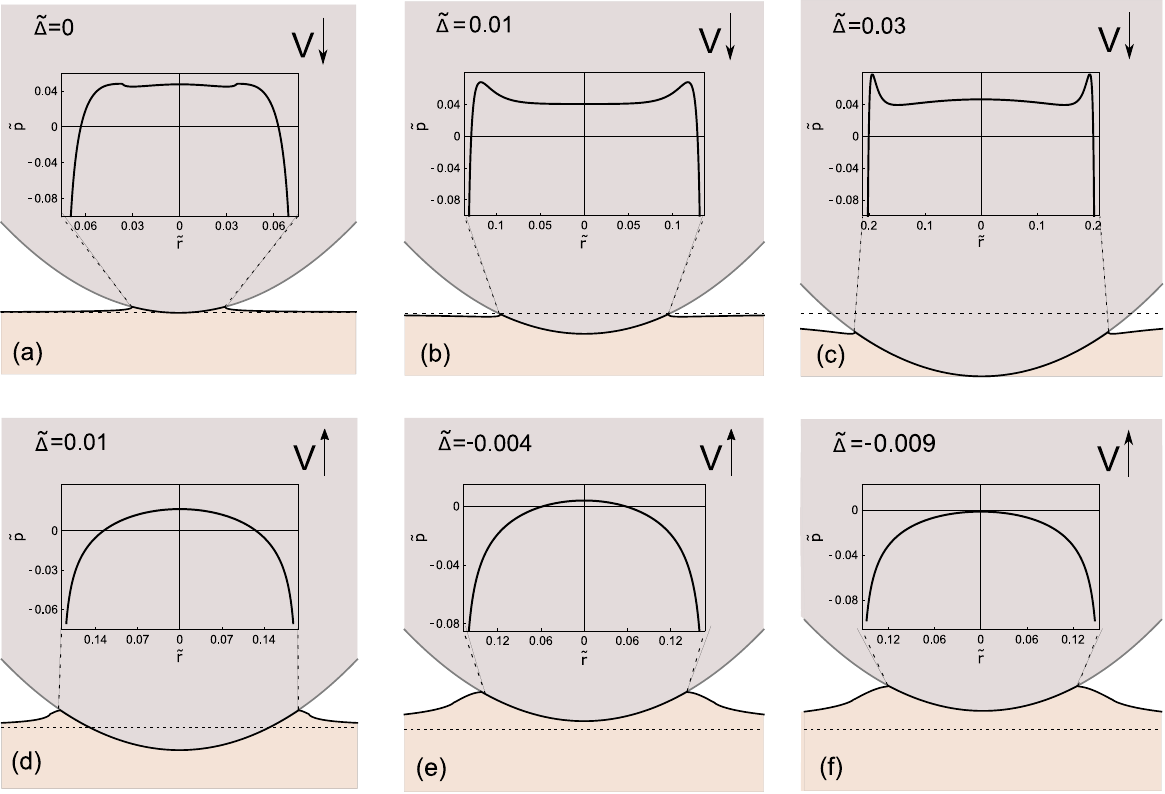} \caption{The deformed contact
configuration during indentation (a-c) and retraction (d-f) at dimensionless
sphere speed $\tilde{V}=0.002$ (i.e, blue line in Fig.\ref{fig2}) for
different values of the dimensionless penetration $\tilde{\Delta}$. The inset
shows the corresponding dimensionless contact pressure distribution $\tilde
{p}$. Results are shown for $\tilde{\gamma}=1.6\times10^{-4},$ $E_{\infty}/E_{0}%
=10$.\newline}%
\label{fig3}%
\end{figure}

Fig. \ref{fig3} reports the surface displacement and the pressure
distributions during an approach-retraction cycle at the given dimensionless
speed $\tilde{V}=0.002$. Focusing on the approach stage, beside the expected
adhesion-induced square root singularity at the contact boundary, the
interfacial pressure distribution also presents a positive annular peak close
to the advancing circular perimeter of the contact area. A similar trend has
been reported for viscoelastic adhesiveless approaching contacts in Ref.
\cite{Putignano2022} and at the leading edge of rolling (or frictionless
sliding) viscoelastic contacts
\cite{Carbone2013,Menga2014,Menga2016visco,Carbone2022}. Moreover, during the
early stages of the retraction process, the size of the contact area
negligibly changes, and it drops only once the maximum tensile load is
reached, as also confirmed by experimental observation
\cite{Violano,Ahn,Li,Lorenz2013}. Noteworthy, during the retraction stage, the
annular pressure peak disappears, and the adhesion-induced pressure
singularity is associated with a trumpet-like opening crack shape, as
predicted by de Gennes \cite{de Gennes}.

\begin{figure}[ptbh]
\includegraphics[width=.98\textwidth]{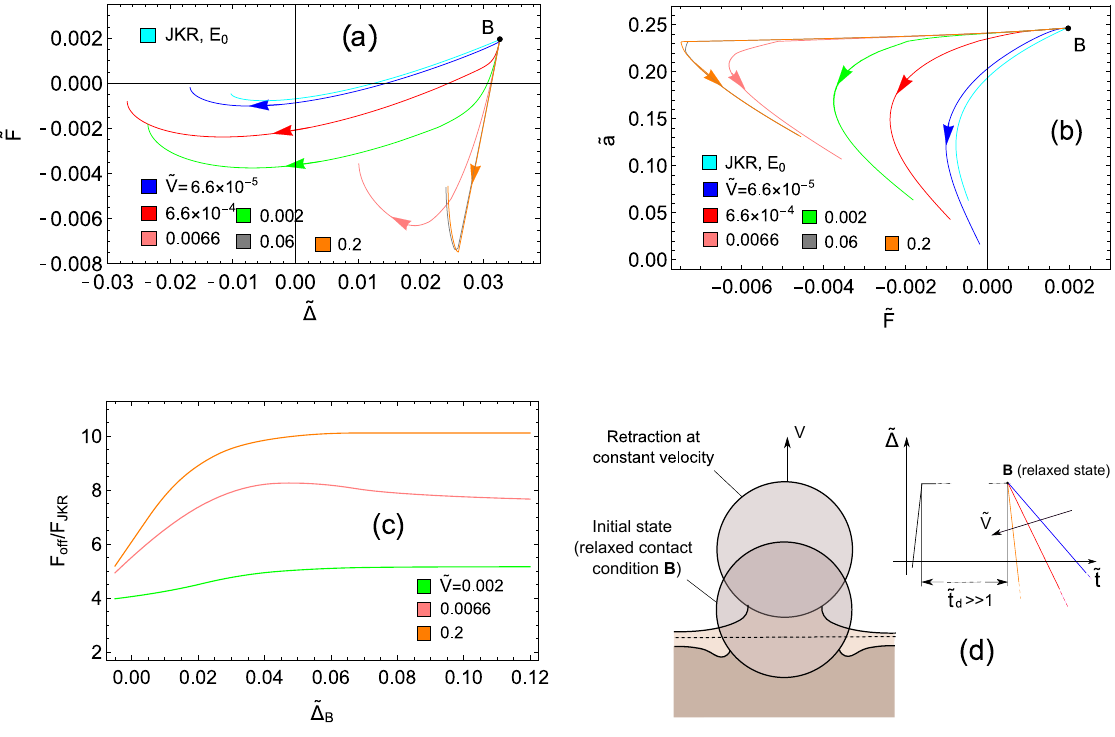}
\caption{Sphere retraction at different dimensionless speed $\tilde{V}$ from
fully relaxed conditions (point B, with $\tilde{\Delta}_{\mathrm{B}}=0.032$).
(a) The dimensionless applied load $\tilde{F}$ vs. the dimensionless
indentation depth $\tilde{\Delta}$. (b) The dimensionless contact radius
$\tilde{a}$ vs. the dimensionless applied load $\tilde{F}$. (c) The normalized
pull-off force $F_{\mathrm{off}}/F_{\mathrm{JKR}}$ vs. the initial
dimensionless penetration $\tilde{\Delta}_{\mathrm{B}}$. (d) The process
schematic and qualitative time-history. Results are shown for $\tilde{\gamma}=1.6\times10^{-4},$
$E_{\infty}/E_{0}=10$.}%
\label{fig4}%
\end{figure}

Results in Fig. \ref{fig2} have shown that a viscoelastic-induced enhancement
of the pull-off force\ can be observed at intermediate approach-retraction
speeds, i.e. when the material has not yet fully entered the glassy state
during the approach stage. Hence, one may guess that the enhancement of
pull-off force or, equivalently, of the adhesion strength should be even more
amplified if the retraction stage is allowed to begin, at finite speed,
immediately after an extremely slow approach stage. This correspond to
retraction from fully relaxed conditions at fixed positive penetration
$\Delta>0$ (dwell time $t_{\mathrm{d}}\rightarrow+\infty$), which is what we
report in Fig. \ref{fig4} for different dimensionless retraction speeds $\tilde{V}$.
Specifically, retraction starts from point B at given dimensionless
penetration $\tilde{\Delta}_{\mathrm{B}}=0.032$ which, as predicted by the
fully relaxed elastic JKR solution (i.e., with elastic modulus $E_{0}$),
corresponds to dimensionless contact radius $\tilde{a}_{\mathrm{B}}=0.245$ and
dimensionless normal load $\tilde{F}_{\mathrm{B}}=1.87\times10^{-3}$. During
retraction, given a finite speed $V$, the contact penetration is
$\Delta=\Delta_{B}-H\left(  t\right)  Vt$, where $H\left(  t\right)  $ is the
unit step function, and the retraction speed instantaneously jumps from $0$ to
$-V$ at $t=0$ (notice, $\dot{\Delta}=-H\left(  t\right)  V$). Therefore, at
the early stages of the retraction process (i.e., for $t\ll\tau$), the
material response is elastic, with modulus $E_{\infty}$. Now, considering
that: (i) a decrease of the contact area can only occur if the condition
$g\left(  r,t\right)  \geq0$ is fulfilled, (ii) a certain time or,
equivalently, a certain retraction distance is required before enough elastic
energy is stored into the system, and (iii) a reduction of the contact area
can only take place if the release of mechanical plus elastic energy is enough
to compensate the change of adhesion energy, one concludes that during the
initial stages of the retraction process the contact area will remain almost
constant in a, say, `frozen' state. When this happens the relation between the
applied load $F$ and the penetration $\Delta$ must obey the flat-punch linear
relation \cite{Maugis1992,Kendall1971}. Force-penetration linearity is indeed
observed in Fig. \ref{fig4} during the initial stages of retraction,
regardless of the given finite speed $V$, and a similar behavior is also
observed in temperature controlled systems \cite{Akulichev2018}, when the
deformed material is cooled below the glass transition temperature
$T_{\mathrm{g}}$ resulting in an almost `frozen' contact shape. Provided that
the material relaxation process has not yet started at detachment (i.e., the
retraction velocity is sufficiently high), the pull-off force $F_{\mathrm{off}%
}$ is much larger that the JKR prediction $F_{\mathrm{JKR}}$, i.e.
$F_{\mathrm{off}}>F_{\mathrm{JKR}}=3\Delta\gamma\pi R/2$. As a consequence, we
conclude that during fast retraction: (i) the material is in the glassy state
hence, att pull-off the energy release rate $G=K_{\mathrm{I}}^{2}/\left(
2E^{*}_{\infty}\right)  $  with $E^{*}_{\infty}=E_{\infty}\left( 1-\nu^{2}\right) $ must be necessarily equal to the adhesion energy per
unit area, i.e. $G=\Delta\gamma$ (the material is not tougher and no
enhancement of the effective energy of adhesion occurs- see also Sec.
\ref{sec:energy}), (ii) the force - penetration and the force - area curves
are significantly different from JKR predictions, (iii) the pull-off force
cannot be predicted by JKR theory. Therefore, any experimental/numerical
estimation of the effective adhesion energy at high speed pull-off through JKR
is inappropriate
\cite{Giri,Greenwood2006,Johnson2000,Afferrante2022,Violano2022,Lorenz2013}.
Conversely, during slow retraction the material has enough time to partially
relax. As a consequence, the maximum tensile force decreases compared to the
value predicted by the aforementioned arguments and monotonically diminishes
with decreasing $V$, eventually reaching the elastic JKR\ value for extremely
slow retraction process. Contact stickiness and toughness are related to the
minimum (negative) value of $\Delta$ before pull-off, as described in
\cite{Menga2016,Menga2019layered,Carbone2022,Mandriota}. In this regard, we
note that for very small values of $V$ the soft elastic JKR\ limit is
recovered, but, at intermediate values of $V$, a significantly larger
elongations before pull-off is observed, compared to the elastic case. This
happens when hysteretic viscoelastic losses occur only close to the circular
boundary of the contact (small-scale viscoelasticity), where the material is
excited on a time scale of order $\rho/\left\vert \dot{a}\right\vert
\approx\rho/V\approx\tau$, where $\rho\ll R$ is the radius of curvature of the
contact adhesive neck. In such conditions, the bulk of the material is instead
excited on time scales of order $R/V>\tau$, thus behaving as a soft elastic
material with modulus $E_{0}$. In such `small-scale viscoelasticity' regime
the energy release rate increases (as better reported in Sec. \ref{sec:energy}%
), and the load-penetration and load-area curves are well approximated by the
JKR\ predictions, provided that the adhesion energy $\Delta\gamma$ is replaced
by an effective value $\Delta\gamma_{\mathrm{eff}}=G$ (where $G$ is the energy
release rate). Note that only in this case the pull-off force can be correctly
estimated by using the adapted JKR theory, i.e.\ $F_{\mathrm{off}}=\left(
F_{\mathrm{JKR}}\right)  _{\mathrm{eff}}=3\pi\Delta\gamma_{\mathrm{eff}}R/2$.
Fig. \ref{fig4} also reports the effect of the initial contact penetration
$\tilde{\Delta}_{\mathrm{B}}$ on the normalized fast retraction pull-off force
$F_{\mathrm{off}}/F_{\mathrm{JKR}}$ and shows that the maximum pull-off
enhancement reaches a plateau at high penetration (i.e., high preload), with
$F_{\mathrm{off}}/F_{\mathrm{JKR}}\approx E_{\infty}/E_{0}$ at very high retraction speed. Reducing the speed leads to smaller values of $F_{\mathrm{off}}/F_{\mathrm{JKR}}$, in agreement with \cite{Violano2022}.
\begin{figure}[ptbh]
\includegraphics[width=.98\textwidth]{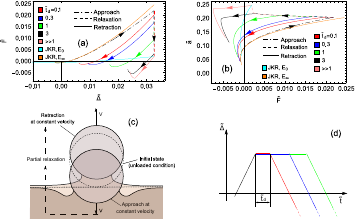}
\caption{Approach-retraction cycles with non-vanishing dimensionless dwell
time $\tilde{t}_{d}$, allowing for partial material relaxation. The
dimensionless sphere speeed is $\tilde{V}=0.1$. (a) The $\tilde{F}$ vs.
$\tilde{\Delta}$ and (b) the $\tilde{a}$ vs. $\tilde{F}$ equilibrium diagrams
for different values of $\tilde{t}_{d}$. (c) and (d) are the process schematic
and the qualitative dimensionless penetration $\tilde{\Delta}$ time-history,
respectively. Results are shown for $\tilde{\gamma}=1.6\times10^{-4},$ $E_{\infty
}/E_{0}=10$.\newline}%
\label{Fig9}%
\end{figure}

In Fig.\ref{Fig9} we report the effect on the adhesive contact behavior of the
dimensionless dwell time $\tilde{t}_{\mathrm{d}}$, i.e. the time delay between
the end of the approaching stage and the initiation of the retraction process.
Since the value $\tilde{t}_{\mathrm{d}}$ physically alters the stress-strain
history, it necessarily affects the system response during retraction, while
the approach stage is unaffected. We present results for dimensionless speed
$\tilde{V}=0.1$, and show that increasing $\tilde{t}_{\mathrm{d}}$ above
$1$\emph{ }yields larger pull-off forces. To understand this peculiar
behavior, we observe that the material relaxation increases with $\tilde
{t}_{\mathrm{d}}$ and, for $\tilde{t}_{\mathrm{d}}\gg1$, the retraction
behavior approaches the flat punch with very large pull off forces, as
discussed above. Moreover, the A-R speed $\tilde{V}$ plays a central role, as
the increasing $\tilde{V}$ require larger $\tilde{t}_{\mathrm{d}}$ values to
achieve relaxation and, in turn, enter the flat-punch regime.

\begin{figure}[ptbh]
\includegraphics[width=.98\textwidth]{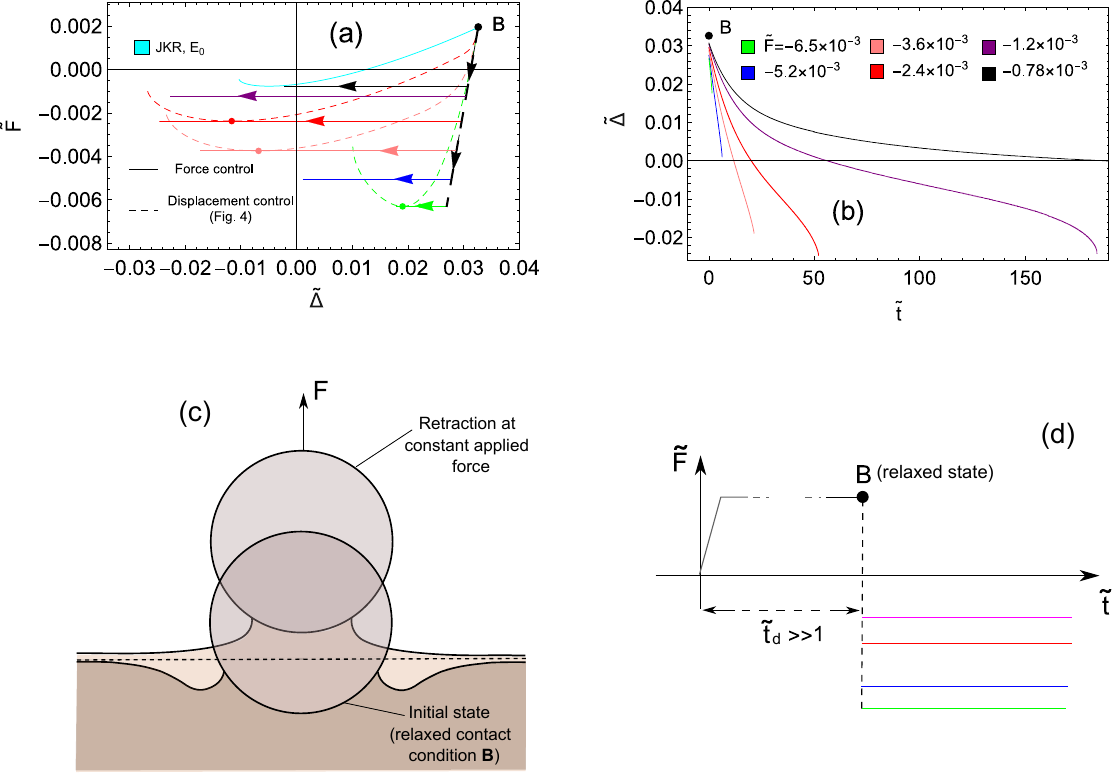}
\caption{The contact behavior under constant tensile force $\tilde{F}$
instantaneously applied once the fully relaxed elastic condition is recovered
(point B, with $\tilde{\Delta}_{\mathrm{B}}=0.032$). (a) The $\tilde{F}$ vs.
$\tilde{\Delta}$ equilibrium diagram for different values of the applied
tensile force $\tilde{F}$ (solid lines); in the same figure, the dashed line
is the behavior at constant retraction velocity corresponding the same
pull-off force. (b) The dimensionless indentation depth $\tilde{\Delta}$ shown
as function of the dimensionless time $\tilde{t}$ for different values of the
applied tensile force $\tilde{F}$. (c) and (d) are the process schematic and
the qualitative dimensionless force $\tilde{F}$ time-history,
respectively. Results are shown for $\tilde{\gamma}=1.6\times10^{-4},$ $E_{\infty
}/E_{0}=10$.\newline}%
\label{fig5}%
\end{figure}

Figs. \ref{fig5} reports the retraction behavior under load-controlled
conditions assuming the same fully relaxed initial conditions (point B in the
figure) as in Figs. \ref{fig4}. This time the tensile force is instantaneously
applied following a step change to the negative value $F_{0}<-F_{\mathrm{JKR}%
}$, i.e. $F\left(  t\right)  =F_{\mathrm{B}}+\left(  F_{0}-F_{\mathrm{B}%
}\right)  H\left(  t\right)  $. The application of this step change in the
applied load leads to an initial high frequency glassy response of the system
with modulus $E_{\infty}$ so that the linear flat punch behavior is again
recovered at the initial stages of the load controlled retraction. Then, the
penetration also jumps to $\Delta_{0}$ following the relation $F_{0}%
=F_{\mathrm{B}}-2a_{\mathrm{B}}E_{\infty}^{\ast}\left(  \Delta_{\mathrm{B}%
}-\Delta_{0}\right)$. After this step change, the contact area and the penetration
$\Delta$ monotonically decrease with time, as well as the retraction speed,
and eventually becomes unstable and detach [see Fig. \ref{fig5}(b)]. This
happens at significantly larger elongations (i.e., larger contact toughness)
compared to $V$-controlled (dashed) curves associated with the same pull-off
forces $F_{0}$.

\begin{figure}[ptbh]
\includegraphics[width=.98\textwidth]{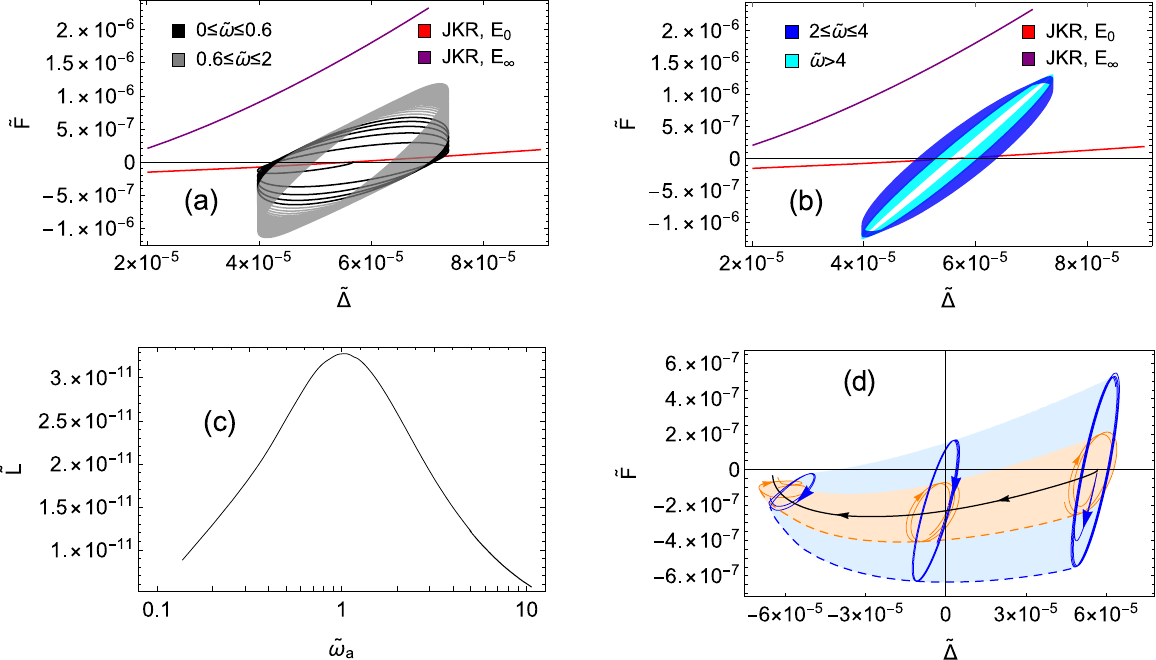}
\caption{Results for normal oscillations: (a,b,c) frequency (up-)sweep around
a given dimensionless penetration $\tilde{\Delta}_{0}$ with $\tilde{\Delta
}\left(  \tilde{t}\right)  =\tilde{\Delta}_{0}+\tilde{\Delta}_{1}\sin\left[
\tilde{\omega}\left(  \tilde{t}\right)  \tilde{t}/2\right]  $ and
$\tilde{\omega}(\tilde{t})=\alpha\tilde{t}$; (d) constant frequency
oscillation superimposed to steady retraction at $\tilde{V}=10^{-7}$ from
$\tilde{\Delta}_{0}$ with $\tilde{\Delta}\left(  \tilde{t}\right)
=\tilde{\Delta}_{0}-\tilde{V}\tilde{t}+\tilde{\Delta}_{2}\sin\left[
\tilde{\omega}\tilde{t}\right]  $. Specifically, (a,b) are the equilibrium
diagram for $\tilde{F}$ vs. $\tilde{\Delta}(t)$, (c) is the dimensionless
energy $\tilde{L}=(1-\nu^{2})L/(\pi E_{0}R^{3})$ dissipated per cycle vs. the
dimensionless frequency $\tilde{\omega}_{a}$ averaged per cycle, and (d) are
the equilibrium diagrams $\tilde{F}$ vs. $\tilde{\Delta}(t)$ for
$\tilde{\omega}=0.35$ (orange line) and $\tilde{\omega}=1.4$ (blue line).
Results refer to $\tilde{\gamma}=5\times10^{-8}$, $E_{\infty}/E_{0}=10$,
$\tilde{\Delta}_{0}=5.7\times10^{-5}$, $\tilde{\Delta}_{1}=-1.7\times10^{-5}$,
$\tilde{\Delta}_{2}=-8.5\times10^{-6}$, and $\alpha=0.006$.}%
\label{fig6}%
\end{figure}

Fig. \ref{fig6} reports the contact behavior of the contact when the
penetration oscillates and refers to $\tilde{\gamma}=5\times10^{-8}$ and
$\delta\tilde{a}=8.5\times10^{-4}$. Specifically, Fig. \ref{fig6}(a,b,c)
presents results for a linear frequency sweep, i.e. $\tilde{\Delta}\left(
\tilde{t}\right)  =\tilde{\Delta}_{0}+\tilde{\Delta}_{1}\sin\left[
\tilde{\omega}\left(  \tilde{t}\right)  \tilde{t}/2\right]  $,\emph{ }with
$\tilde{\omega}=\tau\left(  d\theta/dt\right)  =\alpha\tilde{t}$,
$\alpha=0.006$, $\tilde{\Delta}_{1}=-1.7\times10^{-5}$. The the initial
penetration $\tilde{\Delta}_{0}=5.7\times10^{-5}$ corresponds to the
fully-relaxed (modulus $E_{0}$) JKR solution for $\tilde{F}=0$. As the
frequency is increased [see Fig.\ref{fig6}(a,b)] the slope of the $F$ vs.
$\Delta$ curve increases because of the material stiffening, see also
\cite{Afferrante2022,Menga2016visco}. The amount of energy dissipated per
cycle $\tilde{L}=(1-\nu^{2})L/(\pi E_{0}R^{3})$ roughly equates the area of
the cycle in the $\tilde{\Delta}$ vs. $\tilde{F}$ diagram and is reported in
Fig. \ref{fig6}(c) as a function of the external excitation frequency
$\tilde{\omega}$. The observed bell-shaped behavior is expected, as at very
low and very high excitation frequencies the material behaves elastically,
with vanishing hysteresis. At intermediate frequency (i.e., $\tilde{\omega
}\approx1$) the material response is in the transition region of the
viscoelastic spectrum, and the viscoelastic energy dissipation takes its
maximum value. More interestingly, Fig. \ref{fig6}(d) refers to the case of
constant frequency oscillations (at different frequencies) superimposed to
steady retraction at speed $\tilde{V}=10^{-7}$, so that $\tilde{\Delta}\left(
\tilde{t}\right)  =\tilde{\Delta}_{0}-\tilde{V}\tilde{t}+\tilde{\Delta}%
_{2}\sin\left(  \tilde{\omega}\tilde{t}\right)  $ with $\tilde{\Delta}%
_{2}=-8.5\times10^{-6}$, $\tilde{\omega}=0.35$ (orange line) and
$\tilde{\omega}=1.4$ (blue line). Noteworthy, the envelopes of the maximum
tensile loads (dashed lines) reached during the oscillating retraction stage
significantly increase compared to the case of steady retraction (black curve,
with no oscillations), thus paving the way to possible vibration based
techniques to control interfacial adhesive strength, as experimentally
observed by Shui \textit{et al}. \cite{Shui}.

\subsection{The energy release rate, the elastic energy, and viscoelastic
energy dissipation. \label{sec:energy}}

In viscoelastic materials undergoing deformations, the work of internal
stresses is partially stored as elastic potential energy and partially
dissipated, leading to viscoelastic hysteresis. Neglecting kinetic energy or
inertia forces, energy balance requires the work per unit time of external and
internal forces to be equal, i.e.%
\begin{equation}
F~\dot{\Delta}+\Delta\gamma\dot{A}=\dot{U}+P_{\mathrm{d}}
\label{energybalance}%
\end{equation}
where $\dot{U}$ and $P_{\mathrm{d}}$ are the time-derivative of the stored
elastic energy and the hysteretic energy losses per unit time, respectively.
Most importantly, the energy release rate $G$ can be defined also for non
conservative materials \cite{Carbone2022, Mandriota}, as the change in the
total mechanical energy per unit change in the contact area. Therefore, from
Eq. (\ref{energybalance}), we have
\begin{equation}
G=\frac{\dot{U}}{\dot{A}}-F\frac{\dot{\Delta}}{\dot{A}}=\frac{dU}{dA}%
-F\frac{d\Delta}{dA}, \label{enrelrate}%
\end{equation}
and
\begin{equation}
G=\Delta\gamma-P_{\mathrm{d}}/\dot{A} \label{enrelrate2}%
\end{equation}
which shows that $G$ is a key quantity in adhesive contact mechanics,
sometimes referred to as the effective energy of adhesion or, in other words,
the generalized driving force inducing the contact area change. Consequently,
calculating $G$ is a crucial (and usually tough) task, which requires to
determine either $\dot{U}$ or $P_{\mathrm{d}}$ as functions of the interfacial
stress distribution $\sigma(\mathbf{x},t)$. Aiming at accomplishing this task,
we calculate the work per unit time $P$ done by the internal stresses which,
at equilibrium is only related to the stress and displacement distributions on
the half-space surface. In the most general case (i.e., neglecting axial
symmetry), we have%
\begin{align}
P\left(  t\right)   &  =\int d^{2}x\sigma\left(  \mathbf{x},t\right)  \dot
{u}\left(  \mathbf{x},t\right) \nonumber\\
&  =\int d^{2}xd^{2}x_{1}\mathcal{G}\left(  \mathbf{x-x}_{1}\right)
\sigma\left(  \mathbf{x},t\right)  \dot{\varepsilon}\left(  \mathbf{x}%
_{1},t\right)  \label{power}%
\end{align}
where
\begin{equation}
\varepsilon(\mathbf{x},t)=\int_{-\infty}^{t}dt_{1}J(t-t_{1})\dot{\sigma
}(\mathbf{x},t_{1})=\varepsilon_{0}(\mathbf{x},t)+\sum_{k=1}^{n}%
\varepsilon_{k}(\mathbf{x},t) \label{localdeformation}%
\end{equation}
is an apparent local surface strain, $J(t)=E_{\infty}^{-1}+\sum_{k=1}^{n}%
E_{k}^{-1}\left[  1-\exp\left(  -t/\tau_{k}\right)  \right]  $ is the creep
function for a generic linear viscoelastic material with an arbitrary number
$n$ of relaxation times $\tau_{k}$, and%
\begin{align}
\varepsilon_{0}(\mathbf{x},t)  &  =\frac{\sigma(\mathbf{x},t)}{E_{\infty}%
}\label{first}\\
\varepsilon_{k}(\mathbf{x},t)  &  =\int_{-\infty}^{t}dt_{1}\frac{1}{E_{k}%
}\left[  1-\exp\left(  -\frac{t-t_{1}}{\tau_{k}}\right)  \right]  \dot{\sigma
}(\mathbf{x},t_{1})\nonumber
\end{align}
are, respectively, the elastic contribution to $\varepsilon$ associated with
the high-frequency modulus $E_{\infty}$, and the viscoelastic contributions
associated with each single $k$th Voigt element. We then have%
\begin{equation}
\sigma\left(  \mathbf{x},t\right)  =E_{\infty}\varepsilon_{0}\left(
\mathbf{x},t\right)  =E_{k}\varepsilon_{k}\left(  \mathbf{x},t\right)
+\tau_{k}E_{k}\dot{\varepsilon}_{k}\left(  \mathbf{x},t\right)  ,\qquad
k=1,...,n \label{stresssplit}%
\end{equation}
where $E_{k}\varepsilon_{k}(\mathbf{x},t)$ and $\tau_{k}E_{k}\dot{\varepsilon
}_{k}(\mathbf{x},t)$ represent, respectively, the elastic and viscous stress
components associated with the $k$-th Voigt element. Combining Eqs.
(\ref{power}, \ref{localdeformation}, \ref{first}, \ref{stresssplit}) gives the expression of the
elastic $\dot{U}$ and dissipative $P_{\mathrm{d}}$ contributions to $P$ (i.e.,
$P=\dot{U}+P_{\mathrm{d}}$). In particular,%
\begin{equation}
\dot{U}(t)=\int dx^{2}dx_{1}^{2}\mathcal{G}\left(  \mathbf{x-x}_{1}\right)
\left[  E_{\infty}\varepsilon_{0}\left(  \mathbf{x},t\right)  \dot
{\varepsilon}_{0}\left(  \mathbf{x}_{1},t\right)  +\sum_{k=1}^{n}%
E_{k}\varepsilon_{k}\left(  \mathbf{x},t\right)  \dot{\varepsilon}_{k}\left(
\mathbf{x}_{1},t\right)  \right]  \label{UEd}%
\end{equation}
and
\begin{equation}
P_{\mathrm{d}}(t)=\sum_{k=1}^{n}\int dx^{2}dx_{1}^{2}\mathcal{G}%
(\mathbf{x-x}_{1})\tau_{k}E_{k}\dot{\varepsilon}_{k}\left(  \mathbf{x},t\right)
\dot{\varepsilon}_{k}\left(  \mathbf{x}_{1},t\right)
\label{dissipated energy}%
\end{equation}
Note that in Eq. (\ref{UEd}) the quantity $\mathcal{G}\left(  \mathbf{x}%
\right)  $ is a symmetric function so that it is possible to find the
expression of the elastic energy as%
\begin{equation}
U\left(  t\right)  =\frac{1}{2}\int dx^{2}dx_{1}^{2}\mathcal{G}\left(
\mathbf{x-x}_{1}\right)  \left[  E_{\infty}\varepsilon_{0}\left(
\mathbf{x},t\right)  \varepsilon_{0}\left(  \mathbf{x}_{1},t\right)
+\sum_{k=1}^{n}E_{k}\varepsilon_{k}\left(  \mathbf{x},t\right)  \varepsilon
_{k}\left(  \mathbf{x}_{1},t\right)  \right]  \label{U}%
\end{equation}
The energy release rate $G$ can be calculated using Eqs. (\ref{enrelrate},
\ref{UEd}) at any given time $t$ once solved the elastic problem, i.e. for
known values of $\sigma\left(  \mathbf{x},t\right)  $, $u\left(
\mathbf{x},t\right)  $, and the contact domain $\Omega\left(  t\right)  $.

\begin{figure}[ptbh]
\includegraphics[width=.98\textwidth]{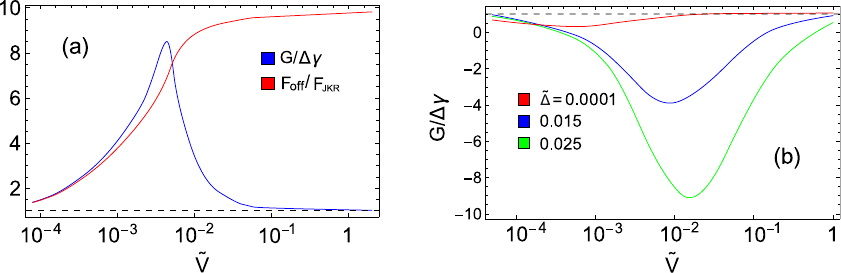}
\caption{ The normalized energy release rate $G/\Delta\gamma$ (a) at pull-off
(blue line) and (b) during indentation (for different dimensionless
penetrations $\tilde{\Delta}$) as functions of the dimensionless
velocity. The viscoelastic-elastic pull-off force ratio
$F_{\text{\textrm{off}}}/F_{\text{\textrm{JKR}}}$ is also shown in (a) as red
line. Initial condition in (a) refers to fully relaxed state (point $B$ in Fig. \ref{fig4}). Results are
shown for $\tilde{\gamma}=1.6\times10^{-4},$ $E_{\infty}/E_{0}=10$.\newline}%
\label{fig7}%
\end{figure}

Recalling that $G=\Delta\gamma-P_{\mathrm{d}}/\dot{A}$ and considering that
$P_{\mathrm{d}}>0$ , we find that during the approach stage (i.e., $\dot{A}%
>0$) the energy release rate $G<\Delta\gamma$ and can be even negative,
whereas during retraction (i.e., $\dot{A}<0$) the energy release rate
$G>\Delta\gamma$. Figure \ref{fig7} reports the trends of the normalized
energy release rate $G/\Delta\gamma$ vs. speed, during (a) retraction and (b)
approach stages. More specifically, $P_{\mathrm{d}}$ vanishes at very-low and
very-large A-R speeds, so that in these two limiting cases $G(t)\rightarrow
\Delta\gamma$. At intermediate speed it increases up to a maximum value, with
the resulting bell-shaped behavior being related to the presence of a
reference length scale (i.e., finite contact size)
\cite{Carbone2022,Mandriota,Persson2021,Persson2017}. For comparison, we
stress that the propagation of a opening semi-infinite crack in an infinite
systems represents a very different scenario, as in this case $P_{\mathrm{d}}$
must monotonically increase with crack speed and eventually reach a plateau
(for isothermal conditions)
\cite{Schapery1989,Persson2005,Greenwood2004,Barber,Carbone2005}. Noteworthy,
since $P_{\mathrm{d}}$ accounts for the energy dissipation in the whole
viscoelastic material, it can also take values such that $P_{\mathrm{d}}%
/\dot{A}>\Delta\gamma$, thus entailing $G(t)<0$ when the sphere is pressed
against the viscoelastic half-space [see Fig. \ref{fig7}(b)]. Fig.
\ref{fig7}(a) also reports the normalized maximum tensile load (i.e. the
pull-off force) as a function of the retraction speed. It is very important to
notice that, despite the bell-shaped rate $G\left(  t\right)  $ vs. $\tilde
{V}$ trend, the ratio $F_{\mathrm{off}}/F_{\mathrm{JKR}}$ continuously
increases until a limiting value is reached ($F_{\mathrm{off}}/F_{\mathrm{JKR}%
}\approx E_{\infty}/E_{0}$). A deeper look at the retraction behavior [Fig.
\ref{fig7}(a)] also shows that $G/\Delta\gamma\approx F_{\mathrm{off}%
}/F_{\mathrm{JKR}}$ only at relatively low retraction speeds, i.e. for
$\tilde{V}\lesssim5\times10^{-4}$. This is the limit where small-scale
viscoelasticity (i.e., localized non-conservative phenomena close to the edge
of the contact) governs the adhesion enhancement
\cite{Carbone2022,Mandriota,Carbone2004}, and JKR pull-off approximation
$F_{\mathrm{off}}=\left(  F_{\mathrm{JKR}}\right)  _{\mathrm{eff}}=3\pi
\Delta\gamma_{\mathrm{eff}}R/2$ holds true (allowing for a rough estimation of
$G=\Delta\gamma_{\mathrm{eff}}$). Differently, at large retraction speeds,
both the small- and large-scale viscoelasticity vanish (e.g., $G\rightarrow
\Delta\gamma$), and the pull-off is governed by the glassy flat punch
behavior, as discussed in Sec. \ref{flat punch}. To the best of authors
knowledge, this is a novel finding, as previous numerical studies on
viscoelastic adhesive contacts (with gap-dependent adhesion)
\cite{Afferrante2022,Violano2022,Lin2002,Mueser2022} relied on JKR pull-off
equation to estimate the energy release rate $G$ even at high retraction speed.

\section{Conclusions}

We study the problem of the unsteady normal indentation of a rigid sphere into
a viscoelastic half-space, in the presence of interfacial adhesion. To solve
the problem, we developed a novel energy-based approach, which generalizes the
Griffith fracture criterion also to time-dependent unsteady non-conservative
contacts. We also present a rigorous procedure to accurately calculate the
time evolution of the elastic energy, the viscoelastic energy dissipation, and
energy release rate $G$, by relying only on the interfacial stress and
displacement distributions. We predict that, depending on the specific
time-history of the contact process, the effective adhesion may be
significantly enhanced by viscoelasticity. At intermediate approach-retraction
speeds, strong adhesive hysteresis is observed because of small-scale
viscoelastic dissipation localized close to the perimeter of the contact area,
which also entails the ability of the system to withstand very high tensile
loads. Hysteresis vanishes at very high and very low approach-retraction speed
as the material response falls, respectively, in the high frequency
(stiff)\ or low frequency (soft) elastic regimes. More importantly, our theory
predicts the extremely large pull-off forces observed experimentally when
retraction starts from a completely relaxed loaded state, with sufficiently
high retraction speed. In this case, the material has no time to relax and
exhibits a `frozen' glassy elastic state, thus resembling the behavior of a
flat-punch with a linear force-penetration relation. We also show that at
sufficiently large retraction speed $V$, the energy release rate reduces with
the increasing $V$ down to the thermodynamic surface energy value
$\Delta\gamma$. However, in such conditions the contact behavior significantly
deviates from JKR theory as small-scale viscoelasticity cannot be invoked in
this case. This implies that the JKR\ model cannot be employed to estimate the
energy release from pull-off force at high retraction speeds, as this
procedure significantly overestimates $G$.

\begin{acknowledgments}
This work was partly supported by the Italian Ministry of University
and Research under the Programme “Department of Excellence
”(decree 232/2016) and partly by the European Union - NextGenerationEU
through the Italian Ministry of University and Research under
the programs: (GC) National Sustainable Mobility Center CN00000023
(decree nr. 1033 - 17/06/2022), Spoke 11 – Innovative Materials
Lightweighting; (NM) PRIN2022 (grant nr. 2022SJ8HTC) and PRIN2022 PNRR (grant nr. P2022MAZHX). The opinions
expressed are those of the authors only and should not be considered as
representative of the European Union or the European Commission's official
position. Neither the European Union nor the European Commission can be held
responsible for them.
\end{acknowledgments}

\appendix

\section{Numerical implementation \label{appA}}

In this section, we outline the numerical procedure to solve Eq.
(\ref{convolution}). We refer to the viscoelastic creep's function with single
relaxation time given by Eq. (\ref{creep2}). In this case, taking the time
derivative of Eq. (\ref{convolution}) leads to
\begin{align}
\dot{u}(\mathbf{x},t)  &  =\frac{1}{E_{\infty}}\int dx_{1}^{2}\mathcal{G}%
(\mathbf{x-x}_{1})\dot{\sigma}(\mathbf{x}_{1},t)\label{displder}\\
&  -\frac{1}{\tau^{2}E_{1}}\exp\left(  -\frac{t}{\tau}\right)  \int_{-\infty
}^{t}dt_{1}\exp\left(  \frac{t_{1}}{\tau}\right)  \int dx_{1}^{2}%
\mathcal{G}(\mathbf{x}-\mathbf{x}_{1})\sigma(\mathbf{x}_{1},t_{1})\\
&  +\frac{1}{\tau E_{1}}\int dx_{1}^{2}\mathcal{G}(\mathbf{x}-\mathbf{x}%
_{1})\sigma(\mathbf{x}_{1},t)\nonumber
\end{align}
Using again Eqs. (\ref{convolution}, \ref{creep2}) yields the following
time-differential equation:%

\begin{equation}
\dot{u}(\mathbf{x},t)=\frac{1}{E_{\infty}}\int dx_{1}^{2}\mathcal{G}%
(\mathbf{x-x}_{1})\dot{\sigma}(\mathbf{x}_{1},t)+\frac{1}{\tau E_{0}}\int
dx_{1}^{2}\mathcal{G}(\mathbf{x-x}_{1})\sigma(\mathbf{x}_{1},t)-\frac
{u(\mathbf{x},t)}{\tau} \label{diffeq}%
\end{equation}

\begin{figure}[ptbh]
\includegraphics[width=.55\textwidth]{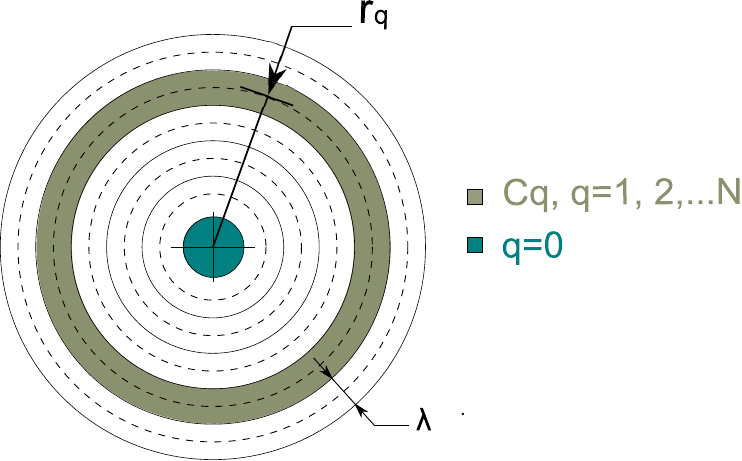}\caption{{The
discretization of spatial domain for the numerical resolution of Eq.
\ref{convolution}}}%
\label{numericalFig}%
\end{figure}which can be solved avoiding integration over the whole time
history. Using a numerical quadrature rule and exploiting the axisymmetric
nature of the problem, the spatial domain is discretized as shown in Fig.
\ref{numericalFig} into $N$ circular annulus $C_{q}$ of constant width
$\lambda$ (with $\lambda/a\ll1$) placed at radii $r_{q}=q\lambda+\lambda/2$,
and a circle $C_{0}$ of radius $\lambda$ placed at $r_{0}=0$. Each area is
subjected to uniform stress $\sigma_{q}\left(  t\right)  =\sigma(r_{q},t),$
$q=0,1,...N$, and Eq.\ref{diffeq} can be rewritten as%
\begin{equation}
\dot{u}_{h}\left(  t\right)  =\dot{u}\left(  r_{h},t\right)  =\frac
{1}{E_{\infty}}\sum_{q=0}^{N}\dot{\sigma}_{q}\left(  t\right)  \int_{C_{q}%
}dx_{1}^{2}\mathcal{G}\left(  \mathbf{x-x}_{1}\right)  +\frac{1}{\tau E_{0}%
}\sum_{q=0}^{N}\sigma_{q}\left(  t\right)  \int_{C_{q}}dx_{1}^{2}%
\mathcal{G}\left(  \mathbf{x-x}_{1}\right)  -\frac{u_{h}\left(  t\right)
}{\tau} \label{discretespace}%
\end{equation}
where $\int_{C_{q}}dx_{1}^{2}\mathcal{G}(\mathbf{x-x}_{1})=\mathcal{\tilde{G}%
}(r_{h},r_{q})=\mathcal{\tilde{G}}_{hq}$ is an axisymmetric field that
represents the surface displacement at radius $r_{h}$ induced on an elastic
half-space of unit modulus by a uniform unit stress acting over $C_{q}$.
$\mathcal{\tilde{G}}_{hq}$ can be easily calculated. Indeed, following
\cite{Johnson,Lubarda2013}, the quantity%

\begin{equation}
\mathcal{K}(r_{h},r_{q})=%
\genfrac{\{}{\vert}{0pt}{}{\left(  1-\nu^{2}\right)  \pi^{-1}4r_{q}%
\mathbf{E(}r_{h}/r_{q}),\text{
\ \ \ \ \ \ \ \ \ \ \ \ \ \ \ \ \ \ \ \ \ \ \ \ \ \ \ \ \ \ \ \ \ \ \ \ \ \ \ \ \ \ \ \ }%
r_{h}\leq r_{q}}{\left(  1-\nu^{2}\right)  \pi^{-1}4r_{h}\left[
\mathbf{E(}r_{q}/r_{h})\text{ }-(1-\left(  r_{q}/r_{h}\right)  ^{2}%
)\mathbf{K(}r_{q}/r_{h})\right]  ,\text{\ \ \ \ }r_{h}>r_{q}}
\label{Kernel}%
\end{equation}
is the surface displacement at radius $r_{h}$ resulting from a uniform unit
stress field acting over a circle of radius $r_{q}$ in the elastic problem
(notably, a similar solution is given in Ref. \cite{Menga2019disp} for uniform
tangential stresses). In Eq. (\ref{Kernel}), $\mathbf{K(}\rho)=\int_{0}%
^{\pi/2}d\xi(1-\rho^{2}\sin^{2}(\xi))^{-1/2}$ and $\mathbf{E(}\rho)=\int
_{0}^{\pi/2}d\xi(1-\rho^{2}\sin^{2}(\xi))^{1/2}$ are the complete elliptic
integrals of the first and second kind, respectively. Thus, according to Fig.
\ref{numericalFig}, $\mathcal{\tilde{G}}_{h0}=\mathcal{K}(r_{h},\lambda)$ and,
using the superposition of effects:%

\begin{equation}
\mathcal{\tilde{G}}_{hq}=\mathcal{K}(r_{h},r_{q}+\lambda/2)-\mathcal{K}%
(r_{h},r_{q}-\lambda/2),\text{ \ \ \ \ }q\neq0 \label{Ghq}%
\end{equation}

The time domain is discretized with small steps $\varepsilon$, with
$\varepsilon/\tau\ll1$ and the discrete form of Eq. \ref{diffeq} can be
written at time $t_{k}=k\varepsilon$ with $k=1,2,.....$.Note that we define
$u_{h}^{k}\left(  r_{h},t_{k}\right)  $ and $\sigma_{h}^{k}=\sigma\left(
r_{h},t_{k}\right)  $ and write%

\begin{equation}
\left(  1+\frac{\varepsilon}{\tau}\right)  u_{h}^{k}=\left(  \frac
{1}{E_{\infty}}+\frac{\varepsilon}{\tau E_{0}}\right)  \sum_{q=0}%
^{N}\mathcal{\tilde{G}}_{hq}\sigma_{q}^{k}+A_{h}^{k-1} \label{linear system}%
\end{equation}
where the quantity%
\begin{equation}
A_{h}^{k}=u_{h}^{k}-\frac{1}{E_{\infty}}\sum_{q=0}^{N}\mathcal{\tilde{G}}%
_{hq}\sigma_{q}^{k} \label{knownterm}%
\end{equation}
\bigskip has already been determined up to time $t^{k-1}$. The linear system
of equations Eq. (\ref{linear system}) allows to calculate, for any given
value of the contact radius $a$, the stress distribution in the contact area
and the displacement and gap distributions out of the contact area. Then,
enforcing the energy balance condition Eq. (\ref{closure}) the equilibrium
value $a^{k}=a\left(  t^{k}\right)  $ of the contact radius can be determined.

\end{document}